\documentclass[prc,twocolumn,epsfig,nofootinbib,floatfix]{revtex4}
\usepackage{graphics}
\usepackage{epsfig}
\usepackage{amsfonts}
\usepackage{amsmath}
\usepackage{bm}

\usepackage{color}

\begin{document}
\title{Toroidal resonance:   relation to
pygmy mode, vortical properties and anomalous deformation splitting}
\author{
V.O. Nesterenko$^{1}$, J. Kvasil$^{2}$, A. Repko$^{2,3}$, W. Kleinig$^{1}$, and P.-G. Reinhard$^{4}$
}
\affiliation{$^{1}$
Laboratory of Theoretical
Physics, Joint Institute for Nuclear Research, Dubna, Moscow region, 141980, Russia}
\email{nester@theor.jinr.ru}
\affiliation{$^{2}$  Institute of
Particle and Nuclear Physics, Charles University, CZ-18000, Prague, Czech Republic}
\affiliation{$^{3}$
Department of Nuclear Physics, Institute of Physics SAS, 84511, Bratislava, Slovakia}
\affiliation{$^{4}$  Institut f\"ur Theoretische Physik II,
Universit\"at Erlangen, D-91058, Erlangen, Germany}

\begin{abstract}
We review a recent progress in investigation of the isoscalar toroidal
dipole resonance (TDR). A possible relation of the TDR and low-energy
dipole strength (also called a pygmy resonance) is analyzed.
It is shown that the dipole strength in the pygmy region can by
understood as a local manifestation of the collective vortical
toroidal motion at the nuclear surface. Application of the TDR as a
measure of the nuclear dipole vorticity is discussed.
Finally, an anomalous splitting of the TDR in deformed nuclei
is scrutinized.
\end{abstract}


\maketitle
\section{Introduction}
\label{intro}

Over the  last years there is a high interest in the
investigation of exotic E1 modes, pygmy dipole resonance (PDR),
compression dipole resonance (CDR), and toroidal dipole resonance
(TDR), see e.g. reviews \cite{Paar07,Savran13}.  The schematic images
of these modes are illustrated in Fig. 1.

PDR is the working title of a collection of marked dipole
strength at energies below the dipole giant resonance. Although the
sub-peaks in the PDR region can have different microscopic
structure, one often visualizes their averaged flow pattern as
oscillations of the neutron excess against the nuclear core with N=Z
\cite{Paar07}. So this resonance can exist only in nuclei with
large neutron excess. As seen from Fig. 1a), the
schematic PDR flow is irrotational. This resonance gathers
effects from various nuclear matter properties (symmetry energy,
incompressibility, effective masses) \cite{Rei13f} and so can provide
useful complementing information on the nuclear equation of state
and thus for astrophysical applications.

The CDR also represents an irrotational flow. It is, as the name
says, a dipole mode, but with compressional character becoming
apparent in a higher order radial profile
\cite{Ha77,St82}, see Fig. 1b). This resonance can be used as a source of
an additional information on the nuclear incompressibility \cite{Paar07}.

Further, the TDR is a remarkable example of {\it vortical} dipole
motion, see Fig. 1c).  For the first time, the toroidal fraction of
the nuclear convection current was inspected by V.M. Dubovik and
A.A. Cheshkov \cite{Dub75}. Later S.F. Semenko predicted the TDR in
atomic nuclei \cite{Se81}.  The TDR is related to CDR
\cite{Paar07,Vr01,Kv11}. In the isoscalar (T=0) channel, both
resonances constitute the low-energy (TDR) and high-energy (CDR) parts
of the isoscalar giant dipole resonance (ISGDR) observed in
$(\alpha,\alpha')$ reaction \cite{Uchida04}. After extraction of the
center-of-mass corrections (cmc), the TDR and CDR become the dominant
dipole modes in the isoscalar channel.

All three exotic modes, PDR, CDR and TDR, have isoscalar (T=0) and
isovector (T=1) branches \cite{Paar07,Savran13,Kv11}.  In the present
review, we will concentrate on the isoscalar branches of the modes.
They are easier accessible in scattering experiments and, as
shown below, deliver indeed rich physics.  Some properties of the
PDR/CDR/TDR were already discussed in the literature
\cite{Paar07,Savran13}. In particular, a modest collectivity of the
modes {were established and it was worked out that complex
  configurations play a crucial role} in formation of the fine
structure of the strengths . In the present review, we will focus on
the vortical representative of this family - isoscalar TDR
\cite{Kv11,Kv13Sn}.  Namely, we will outline some fascinating features
of this TDR, recently explored by our group: relation to the PDR
\cite{Repko13,Dresden15}, possibility to use TDR as a measure of the
dipole nuclear vorticity \cite{Rein14vor}, and anomalous deformation
splitting of the TDR \cite{Kv14def}.  The analysis is based on the
self-consistent calculations within the Skyrme Quasiparticle
Random-Phase-Approximation (QRPA) approach \cite{Skyrme,Vau72,Ben03}
with the factorized \cite{Ne02,Ne06} and exact \cite{Repko} residual
interaction.
\begin{figure}
\includegraphics[width=8cm]{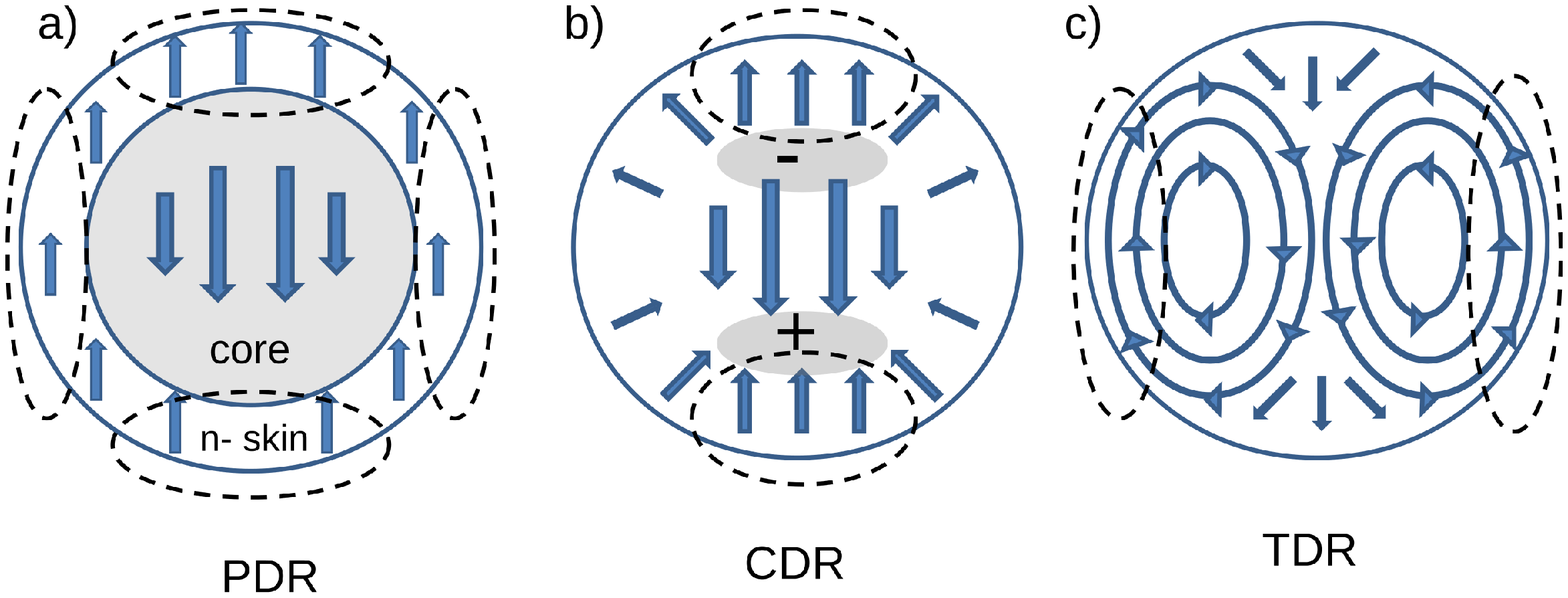}\label{fig1_image}
\caption{Schematic velocity fields for the PDR (a), high-energy CDR (b),
and TDR (c) flows. The driving field is directed along z-axis.
The arrows indicate only directions of the flows but not their strength. In the plot (b),
the compression (+) and decompression (-) regions, characterized by increased and decreased
density, are marked. The dash curves represent the surface regions where the PDR and CDR/TDR
flows are similar.}
\end{figure}

The paper is organized as follows. In Section 2, the basic formalism
for the TDR and CDR is given. The calculation scheme is outlined. In
Section 3, we demonstrate that the PDR can be treated as a
manifestation of the TDR at the nuclear surface. In Section 4, the
possibility to use the toroidal strength as a measure of the nuclear
dipole vorticity is discussed. It is shown that this prescription is
more accurate and robust than the familiar Ravenhall-Wambach recipe
\cite{Ra87}. In Section 5, we inspect the anomalous deformation
splitting of the TDR in axial nuclei (an opposite order of K-branches
as compared to the case of the giant dipole resonance - GDR).  This
feature can be used as an unambiguous fingerprint of the TDR in
experiment.  Finally, in Section 6 the summary is done.

\section{Basic formalism and calculation scheme}
\subsection{Multipole toroidal and compression operators}

In the long-wave approximation ($k \to 0$),
the standard electric multipole operator reads
\begin{equation}\label{E+ktor}
\hat{M}(E\lambda\mu, k) \approx \hat{M}(E\lambda\mu)
+ k\:\hat{M}_{\text{tor}}(E\lambda\mu)
\end{equation}
where
\begin{eqnarray}\label{E_oper_main}
\nonumber
\hat{M}(E\lambda \mu) &=&
 -\frac{i}{kc} \int d^3r \: (\mathbf{\nabla}
\cdot \hat{\mathbf{j}}_{\text{nuc}})
 r^{\lambda} Y_{\lambda \mu}
\\
 &=& - \int d^{3}r \:\hat{\rho}\:
 r^{\lambda} Y_{\lambda\mu}
\end{eqnarray}
is the familiar electric operator  and
\begin{eqnarray}\label{tor_rel_1}
\hat{M}_{\text{tor}} (E\lambda \mu)
&=& \frac{i}{2c(\lambda+1)(2\lambda +3)}
\\
&&\cdot\int d^3r \hat{\mathbf{j}}_{\text{nuc}} \cdot [ \:\mathbf{\nabla}
\times \:(\mathbf{r} \times \mathbf{\nabla}) r^{\lambda +2} Y_{\lambda \mu}]
\nonumber
\\
\nonumber
  &=&
 -\frac{1}{2c}
 \sqrt{\frac{\lambda}{\lambda+1}}\,\frac{1}{2\lambda+3}
 \int d^3r \: r^{\lambda+2}\mathbf{Y}_{\lambda\lambda\mu}
\\ \label{TM_curlj}
 &&\cdot
  \left(\mathbf{\nabla} \times \hat{\mathbf{j}}_{\text{nuc}}\right)
\end{eqnarray}
is the toroidal operator \cite{Dub75,Se81,Kv11}.  Here
$\hat{\mathbf{j}}_{\rm{nuc}}(\mathbf{r})=
\hat{\mathbf{j}}_{\rm{c}}(\mathbf{r})+\hat{\mathbf{j}}_{\rm{m}}(\mathbf{r})$
is the operator of the nuclear current consisting of convection and
magnetization parts. Further, $\hat{\rho}$ is the density operator,
$\mathbf{Y}_{\lambda\lambda\mu}(\hat{\bf r})$ and
$Y_{\lambda\mu}(\hat{\bf r})$ are vector and ordinary spherical
harmonics.  For the sake of brevity, we will skip below (up to the
cases of a possible confusion) the coordinate dependence in currents,
densities and spherical harmonics.

Equation (\ref{E+ktor}) shows that the toroidal
operator comes as a second order ($\sim k^2$) correction to the
dominant electric operator (\ref{E_oper_main}). Being determined by the curl
of the nuclear current $\left(\mathbf{\nabla} \times \hat{\bf j}_{\text c}\right)$, the
toroidal flow is obviously vortical and thus  decoupled
from the continuity equation. So the toroidal operator
cannot be presented through the nuclear density alone and needs
the knowledge of the current distribution.

The multipole compression operator reads \cite{Kv11,Ha77,St82}
\begin{eqnarray}\label{CM_divj}
  \hat{M}_{\text{CM}}(E\lambda\mu)
  &=&
  -
\frac{i}{2c(2\lambda+3)}
 \int d^3r r^{\lambda+2}{Y}_{\lambda\mu}
 \nonumber
 \\
  &\cdot& \left(\mathbf{\nabla}\cdot\hat{\bf j}_{\text{nuc}}\right)
\\
&=&
-k\frac{1}{2(2\lambda+3)}\int d^3r
\hat{\rho} r^{\lambda+2} Y_{\lambda\mu} \;
\label{CM_divj1}
\\
&=&
-k \hat{M}'_{\text{CM}}(E\lambda\mu) \;,
\end{eqnarray}
where $\hat{M}'_{\text{CM}}(E\lambda\mu)$ is its familiar
density-dependent form \cite{Ha77,St82}. The compression operator
appears as a probe operator for excitation of the
ISGDR \cite{Ha77,St82}. Unlike the toroidal one, this
operator may be presented in both current- and density-dependent
forms. The operator is determined by the divergence of current
$\left(\mathbf{\nabla} \cdot \hat{\bf j}_{\text{c}} \right)$
and so is irrotational.

Note that, though the toroidal and compression operators represent
essentially different flows, they are closely
related. Accordingly there is a coupling between the TDR and CDR
\cite{Kv11}.

\subsection{E1(T=0) case}

Now we will focus on the isoscalar dipole
E1(T=0) excitations. Following \cite{Kv11}, the role of the magnetization nuclear
current for TDR and CDR in E1(T=0) channel is negligible.
So only the  convection part  $\hat{\mathbf{j}}_{\text
c}$ of the nuclear current is considered below.

In E1(T=0) channel, the toroidal and compression
operators are reduced to \cite{Kv11,Kv13Sn,Repko13,Rein14vor,Kv14def}
\begin{eqnarray}
\label{TM_oper} && \hat M_{\text{tor}}(E1\mu) = -\frac{i}{2\sqrt{3}c} \int
d^3r \hat{\bf j}_{c}
\\ \nonumber
&&\quad  \cdot \; [\frac{\sqrt{2}}{5}r^2{\mathbf{Y}}_{12\mu} + (r^2
-\langle r^2\rangle_0) {\mathbf{Y}}_{10\mu}] ,
\\
\label{CM_oper}
&& \hat M_{\text{com}}(E1\mu) = -\frac{i}{2\sqrt{3}c} \int
d^3r \hat{\bf j}_{c}
\\ \nonumber
&&\quad  \cdot \; [\frac{2\sqrt{2}}{5} r^2{\mathbf{Y}}_{12\mu} - (r^2 -
\langle r^2\rangle_0) {\mathbf{Y}}_{10\mu}] ,
\\
\label{CMp_oper} && \hat M'_{\text{com}}(E1\mu) =  \frac{1}{10}
\int d^3r \hat{\rho} \; [r^3-\frac{5}{3}\langle r^2\rangle_0 r]
Y_{1\mu} \; .
\end{eqnarray}
Here $\langle r^2 \rangle_0^{\mbox{}}=\int d^3r \rho_0 r^2/A$ is
the ground-state squared radius, $\rho_0({\bf r})$ is the ground
state density, $A$ is the mass number. The operators
(\ref{TM_oper})-(\ref{CMp_oper}) have  the center of mass
correction (c.m.c.) proportional to  $\langle r^2\rangle_0$.

The toroidal and compression matrix elements for $E1$ transitions between
the ground state $|0\rangle$ and RPA excited state $|\nu\rangle$
are determined through the current transition density (we
skip below the index $\mu$)
\begin{equation}\label{CTD}
\delta {\mathbf{j}}_{\nu}({\mathbf{r}}) = \langle \nu| \:
\hat{\mathbf{j}}_{\text c}({\mathbf{r}}) \:| 0 \rangle = -i
[j^{\nu}_{10}(r)\:{\mathbf{Y}}_{10}^* + j^{\nu}_{12}(r) \:{\mathbf{Y}}_{12}^*]
\end{equation}
involving the upper $j_+=j_{12}$ and lower $j_-=j_{10}$ radial current components.
Then
\begin{eqnarray}
\label{TM_me}
 \langle \nu |\hat M_{\text{tor}}(E1)|0\rangle &=& -\frac{1}{6c} \int
dr r^2
\\ \nonumber
&\cdot&  [\frac{\sqrt{2}}{5}r^2 j_{12}^{\nu}
+ (r^2 -\langle r^2\rangle_0) j_{10}^{\nu}] \; ,
\\
\label{CM_me}
 \langle \nu |\hat M_{\text{com}}(E1)|0\rangle &=& -\frac{1}{6c} \int
dr r^2
\\ \nonumber
&\cdot&  [\frac{2\sqrt{2}}{5}r^2 j_{12}^{\nu}
- (r^2 -\langle r^2\rangle_0) j_{10}^{\nu}] \; .
\end{eqnarray}
Ravenhall-Wambach proposed that just the current component $j_+$
determines the vorticity of the dipole motion \cite{Ra87}.
Then the flow can be treated as  fully
vortical ($j_+\ne$0, $j_-$=0), fully irrotational ($j_+=$0,
$j_-\ne$0), and mixed ($j_+\ne$0, $j_-\ne$0). Following this
prescription, both TDR and CDR are of a mixed
(irrotational/vortical) character, which contradicts their
predominantly  curl- and gradient-like velocities fields exhibited in
Fig. 1, see also detailed discussion in \cite{Kv11,Rein14vor}.

\subsection{Details of the calculation}

The calculations were performed within the fully self-consistent Skyrme QRPA method
\cite{Ri80}. We use the exact residual interaction for spherical nuclei \cite{Repko}
and the separable residual interaction  for axial deformed nuclei \cite{Ne06}.
Below these schemes are referred as RPA and SRPA, respectively. Both versions are
self-consistent because: i) the mean field and
residual interaction are obtained from the same Skyrme functional, ii)
the residual interaction includes all the terms of the initial Skyrme
functional as well as the Coulomb direct and exchange terms. Both time-even
and time-odd densities are take into account. The $\delta$-force volume pairing
is treated at the BCS level \cite{Ben00}. The Skyrme force SLy6 \cite{SLy6}
providing a satisfactory description of the giant dipole resonance (GDR) \cite{Kl08} is used.

For a deformed nucleus, the equilibrium quadrupole axial deformation
$\beta$ is determined by minimization of the total energy of the
system. The SRPA code employs a mesh in cylindrical coordinates.

The calculations use a large configuration space with
particle-hole (two-quasiparticle) energies up to 70 MeV.
The energy-weighted sum rule for E1(T=1) transitions
is fully exhausted. The spurious mode lies around 1 MeV,
i.e. safely beyond the TDR and CDR regions.

The toroidal and compression E1(T=0) strength functions read
\begin{equation}\label{SF_TDR}
  S_{\alpha}(E1, T=0) = \sum_{\mu, \nu}
  |\langle\nu|\hat{M}_{\alpha}(E1\mu)|0\rangle|^2
  \zeta(E - E_{\nu})
\end{equation}
where $\hat{M}_{\alpha}(E1\mu)$ is the toroidal ($\alpha=\mathrm{tor}$)
or compression ($\alpha=\mathrm{com}$)
transition operator given by expressions (\ref{TM_oper}) and (\ref{CM_oper}),
respectively. The neutron and proton effective charges are
$e_n^{\mathrm{eff}}=e_p^{\mathrm{eff}}=1$. The sum in (\ref{SF_TDR}) runs through all the
QRPA $\nu$-states. Further, $\zeta(E - E_{\nu})$ is a Lorenz weight with the averaging parameter
$\Delta$. The folding by the Lorentz function simulates the smoothing effects
beyond QRPA (coupling to complex configurations and escape widths).
In SRPA calculation for axial deformed nuclei, we use the constant averaging parameter
$\Delta$=1 MeV \cite{Ne06}. This suffices to demonstrate the deformation
splitting of the TDR, discussed in Sec. 5. However
our RPA code for spherical nuclei  allows to use a more reasonable Lorentz double
folding exhibiting a linear dependence of $\Delta$
on the excitation energy $E$ above the first emission threshold \cite{IJMPE11,Repko}.
This folding is implemented in strength distributions discussed in Sec. 3.

For the description of the GDR and PDR, we
calculate within the RPA the photoabsorption $\sigma_{\gamma}(E1, T=1)$ \cite{BM74}.
The neutron and proton effective charges $e_n^{\mathrm{eff}}=-Z/A$
and $e_p^{\mathrm{eff}}=N/A$ are used. The Lorentz energy dependent
double folding \cite{Repko} is applied.

More details on the calculation scheme can be found elsewhere
\cite{Kv11,Kv13Sn,Repko13,Rein14vor,Kv14def}.

\section{Relation of the toroidal and pygmy modes}

In the studies \cite{Repko13,Dresden15}, it was reported that the PDR
and TDR occupy the same energy region and thus are related. This
point was thoroughly investigated for a doubly magic nucleus
$^{208}$Pb. The strength functions, transition densities and current
fields were inspected. The calculations confirmed the experimental
finding \cite{End10} that the PDR is separated into isoscalar
lower-energy and mixed (isoscalar/isovector) higher-energy
branches.  The main attention was paid to the low-energy isoscalar
part of the PDR \cite{Repko13,Dresden15}.  It was shown that though
the transition densities in this energy region are typical for the PDR
(dominance of the neutron flow at the nuclear surface), the
corresponding current fields (current transition densities) are
obviously of toroidal character. Thus it was suggested that the PDR is
a peripheral manifestation of the toroidal flow.

At first glance, this suggestion looks incorrect because PDR and TDR
represent essentially different irrotational and vortical
kinds of nuclear motion. This is clearly seen in the
panels a) and c) of Fig. 1. However the same panels indicate that both motions are quite similar
at the nuclear surface, see left and right boundary regions marked by the dash curves. So the PDR could be
a local peripheral manifestation of the TDR. Following Fig. 1, PDR and CDR flows are similar at the top
and bottom boundary regions. So perhaps the PDR has also
some dipole compression fraction.
\begin{figure}
\includegraphics[width=\linewidth]{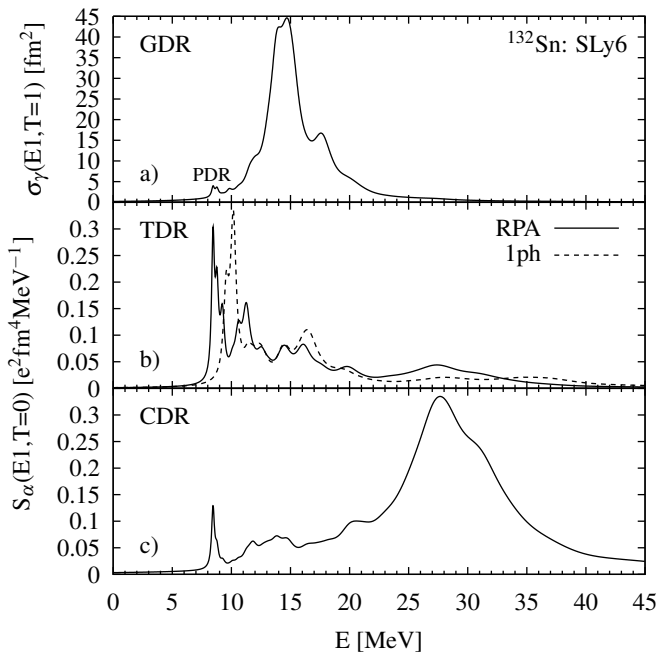}\label{fig2_sf}
\caption{ The dipole strengths calculated within RPA with the force SLy6 in in $^{132}$Sn:
the photoabsorption embracing the GDR and PDR (a); the E1(T=0) strength functions (\ref{SF_TDR})
for the TDR (b)and CDR (c). The panel (b) also shows the 1ph strength
depicted by the dash line.
}
\end{figure}

In this section, we briefly repeat the above analysis but now for a
doubly-magic nucleus $^{132}$Sn characterized by a significant neutron
excess. Let's first consider the relevant strength functions exhibited
in Fig. 2. In the panel a) the photoabsorption $\sigma_{\gamma}$
embracing the GDR and PDR is shown.  The toroidal and compression
strengths (\ref{SF_TDR}) are given in the panels b)-c).  It is easy to
see that PDR and low-energy TDR/CDR peaked parts share the same energy
region and so can be coupled. The panel b) also shows the
particle-hole (1ph) unperturbed toroidal strength.  One sees that, in
the RPA case, the low-energy TDR peak is significantly downshifted
relative to its 1ph counterpart. This signals a noticeable
collectivity of the TDR.

More detailed information of the dipole strength in this energy region
can be obtained from the inspection of the transition densities (TD)
and current fields (CF).  Note that TD and CF do not depend on the
toroidal (\ref{TM_oper}) and compression (\ref{CM_oper}) operators and
are fully determined by the RPA wave functions. It is not worth to
consider these patterns for individual $\nu$-states because they can
vary from state to state and so hide common features of the
flow. Instead we consider TD and CF properly averaged (summed) over a
relevant energy intervals. The technique of computing of the summed
TD/CF is described in detail in \cite{Repko13}. Note that this
technique circumvents a possible ambiguity in the signs of the
RPA wave functions. In the present study for $^{132}$Sn, the TD/CF are
calculated at the energy interval 6-10 MeV covering the PDR and
low-energy TDR and CDR peaks.
\begin{figure}
\includegraphics[width=\linewidth]{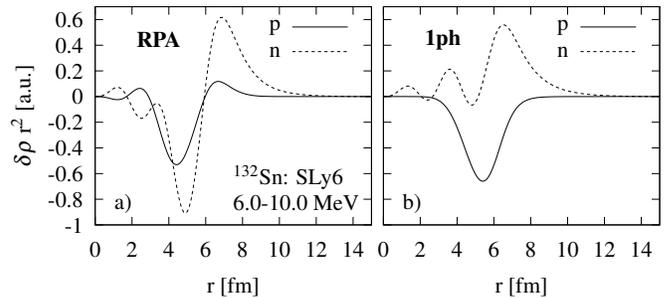}\label{fig3_td}
\caption{RPA (left) and 1ph (right) $r^2$-weighted proton
(solid curve) and neutron (dash curve) transition densities
in $^{132}$Sn, calculated with the force SLy6. The TD are
summed in the energy interval 6-10 MeV.}
\end{figure}

Figure 3 shows the RPA and 1ph proton and neutron transition
densities in $^{132}$Sn. In the RPA case, we see a typical picture
when the proton and neutron TD are strong and in phase inside the
nucleus ($r\sim$4-6 fm) but the neutron TD strictly dominates at the
nuclear boundary ($r\sim$6-10 fm). Just this TD behavior is often
used as a justification of the PDR view in terms of the oscillation of
the neutron excess against the N=Z core, see Fig. 1a). A noticeable
difference between the RPA and 1ph TD indicates that the impact of the
residual interaction is important.
\begin{figure}
\includegraphics[width=\linewidth]{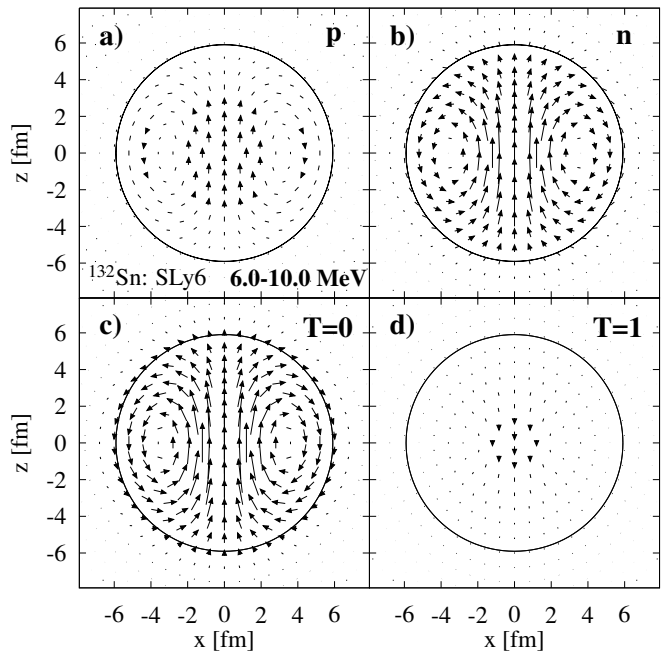}
\caption{The proton (a), neutron (b), isoscalar (c) and isovector (d) current fields
calculated within RPA in $^{132}$Sn. The CF are
summed in the energy interval 6-10 MeV.}
\end{figure}

The TD are still too rough characteristics since they illustrate only
a radial dependence of the flow but not its angular distribution. At
the same time, the latter is crucial to recognize the actual flow
pattern. Thus we have to inspect the CF which reveal more details. The
CF for the energy interval 6-10 MeV in $^{132}$Sn are exhibited in
Fig. 4. The proton, neutron, isoscalar, and isovector flows at the
plane $z-x$ are shown. We see that the nuclear motion at 6-10 MeV has
a clear isoscalar toroidal character, cf. with Fig. 1c).
So it is quite possible that the PDR is rooted in the
TDR and is actually its manifestation at the nuclear surface (where
both PDR and TDR flows look as similar irrotational patterns)
if the surface is dominated by neutron density.
Note that by definition the PDR
exists only in nuclei with a neutron excess. Instead the TDR is a
general feature of all the nuclei including those with N=Z.

The total PDR, with  its isoscalar and mixed parts, is probably a complicated
mixture of different kinds of the dipole motion: toroidal, compression,
GDR tail, etc \cite{Repko13,Dresden15}. However our analysis of CF
obviously shows that the toroidal contribution is of a prior importance.
All the dipole motions should be coupled and so generated simultaneously in
various reactions. A ratio between their contributions in the dipole response
should depend on the applied reaction (photoabsorption, $(\alpha,\alpha')$, $(e,e')$, etc).

Note that some signs of the toroidal flow in the PDR region can be
noticed in the velocity fields of previous RPA calculations
\cite{Vr01,De14}. A distinctive toroidal motion in the PDR region was
obtained in the calculations \cite{Ry02} within the Quasiparticle
Phonon Model (QPM) \cite{Sol76}. Besides, the PDR/TDR interplay was
recently discussed in the semiclassical exploration
\cite{Ur12}. However the studies \cite{Vr01,De14,Ry02} did not
consider a possible relation between PDR and TDR while the study
\cite{Ur12} was done at a phenomenological level. Besides, unlike our
scheme, the previous RPA and QPM calculations inspected TD/CF for
individual states or used summed TD/CF without a special care for an
ambiguity in the sign of the dipole states.

\section{TDR as a measure of nuclear dipole vorticity}

There are two basic kinds of the nuclear flow: irrotational and vortical \cite{Ri80,BM74}.
The irrotational motion is pertinent to collective low-energy excitations
and regular electric GR \cite{Harakeh_book_01}. The vortical flow takes place
(besides a plane nuclear rotation) in single-particle excitations
\cite{Ra87} and exotic GR like e.g. electric TDR \cite{Kv11,Rein14vor} and twist
magnetic quadrupole resonance \cite{HE_77}).

Despite some previous studies (see e.g. \cite{Ra87,Ca99}), our knowledge
about nuclear vorticity is still rather poor. Even the measure of the
vorticity is disputable. In hydrodynamics,
the vorticity is defined as curl of the velocity field \cite{Lan87},
\begin{equation}\label{HDV}
  \bf \varpi (\mathbf{r})=
\mathbf{\nabla} \times \mathbf{v} (\mathbf{r}) \; .
\end{equation}
However nuclear physics deals not with velocities but
currents. In this connection, Ravenhall and Wambach have proposed
the $j_{+}$-component of the nuclear current as a  measure and indicator
of the nuclear vortcity (to be called below as RW vorticity) \cite{Ra87}.
They have shown that $j_{+}$ is unrestricted by the continuity equation (CE)
\begin{equation}\label{ce_ctd}
\delta\dot{\rho}_{\nu} ({\mathbf{r}}) + {\mathbf{\nabla}}
\cdot \delta{\mathbf{j}}_{\nu}({\mathbf{r}}) = 0
\end{equation}
in the integral sense and so is suitable for a divergence-free
(vortical) pattern \cite{Ra87}.

For a long time, the RW prescription was used for estimation of the
vortical contribution in various excitations, see
e.g. \cite{Ry02}. However our last studies have shown that this
prescription has serious shortcomings \cite{Kv11,Rein14vor}.  In
particular, the RW scheme obviously fails for the TDR and CDR. Indeed,
following Eqs. (\ref{TM_me})-(\ref{CM_me}), these modes involve both
$j_{+}$ and $j_{-}$ components of the nuclear current and so have to
be of a mixed (irrotational/vortical) nature. At the same time, the
TDR is basically vortical while the CDR is irrotational.  Their
velocities are the gradient and curl functions, respectively
\cite{Kv11}.  And their current fields closely correspond to vortical
and irrotational images given in Fig. 1.

Moreover, the thorough analysis of the current components
$j_{-}({\mathbf{r}})=j^{\nu}_{10}(r) \:{\mathbf{Y}}_{10}^*$ and
$j_{+}({\mathbf{r}})=j^{\nu}_{12}(r) \:{\mathbf{Y}}_{12}^*$ has shown
that both them have strong contributions in the low-energy toroidal
and high-energy compression regions \cite{Rein14vor}.  Besides, the
divergences and curls of these components turned out to be of the
same order of magnitude. This means that $j_{+}$ has no any strong
advantage over $j_{-}$ to represent the nuclear vorticity. Neither
$j_{+}$ nor $j_{-}$ alone can represent a vortical or irrotational
flow. Instead Eqs. (\ref{TM_me}) and (\ref{CM_me}) suggest that only
the proper combinations of $j_{+}$ are $j_{-}$ are suitable for these
aims.

In this connection, we propose the toroidal strength as a natural and robust
measure of the nuclear vorticity. Following Eqs. (\ref{TM_me}), the toroidal
matrix element is determined by the combination of $\delta j_{+}$ and
$\delta j_{-}$, resulting in a clear  vortical flow demonstrated in Figs. 4.
Besides, as shown in \cite{Rein14vor}, the toroidal and hydrodynamical
criteria of the vorticity are closely related.

One may also provide some formal arguments to justify the toroidal
pattern as the standard for vorticity. We know that the electric current
transition density is decomposed into longitudinal and transversal
components,
\begin{eqnarray}
  \delta \mathbf{j}(\mathbf{r})&=&\delta \mathbf{j}_{\parallel}(\mathbf{r})
   + \delta \mathbf{j}_{\perp}(\mathbf{r}) \:,
   \\
\delta \mathbf{j}_{\parallel}(\mathbf{r}) &=& \mathbf{\nabla}
  \phi(\mathbf{r}), \quad
  \delta \mathbf{j}_{\perp}(\mathbf{r}) = \mathbf{\nabla}
  \times \mathbf{\nabla} \times (\mathbf{r}\chi(\mathbf{r})) ,
\label{cur_LT}
\end{eqnarray}
where $\phi(\mathbf{r})$ and $\chi(\mathbf{r})$ are some scalar
functions \cite{Dub75}.  As compared to the prescription \cite{Ra87},
this definition looks more plausible for the search of
CE-unrestricted divergence-free current because we immediately obtain
$\delta \mathbf{j}_{\perp}$ as a natural candidate.

The current components can be expanded in the basis of eigenfunctions
$\it{\mathbf{J}}^{(\kappa)}_{\lambda\mu k}({\mathbf{r}})$ ($\kappa$ =
-, 0, +) of the vector Helmholtz equation (in analogy to the expansion
of the vector-potential, in \cite{EG_book}). Then the transversal
component reads
\begin{equation}
\delta \mathbf{j}_{\perp}(\mathbf{r})=\sum_{\lambda \mu k}\it{\mathbf{J}}^{(+)}_{\lambda\mu k}
({\bf r}) \; m^{(+)}_{\lambda\mu} (k)
\end{equation}
where $m^{(+)}_{\lambda\mu} (k)$ is the electric transversal
form-factor and integration by $k$ is assumed. In the long-wave
approximation ($k \to 0$), the transversal component is reduced to the
longitudinal one. After subtraction of this long-wave part from $\delta
\mathbf{j}_{\perp}$, we get at $k>0$ the toroidal current density
\cite{Dub75}.  The transversal character of such toroidal current is
confirmed by expression (\ref{tor_rel_1}).  Being independent from
$\delta\mathbf{j}_{\parallel}$ and thus decoupled from CE, the
toroidal current is a relevant vortical part of the complete nuclear
current, suitable as a robust measure of the nuclear
vorticity.

\section{Anomalous deformation splitting of TDR}

The GDR in axial nuclei exhibits the deformation splitting into K=0
and K=1 branches \cite{Da58}.  In prolate nuclei, the branch K=0 has a
lower energy than the K=1 one, $E_{K=0}<E_{K=1}$.  The opposite
sequence, $E_{K=1}<E_{K=0}$, takes place in oblate nuclei. This
splitting can be easily explained in terms of dipole oscillations
along $z$- and $x,y$-axes.

The deformation splitting should also take place in the TDR. However,
the calculations for the TDR (prolate $^{170}$Yb \cite{Kv14def}) have
surprisingly revealed an anomalous sequence $E_{K=1} < E_{K=0}$,
i.e. opposite to the order typical for GDR in oblate axial
nuclei. It was shown that this sequence takes place in both SRPA and
2qp toroidal strength functions. So the effect is not caused by the
residual interaction.
\begin{figure}
\includegraphics[width=\linewidth]{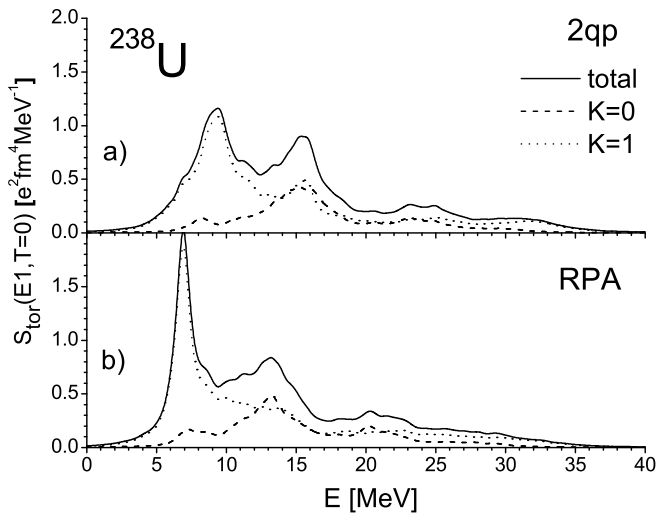}
\caption{2qp (a) and SRPA (b) TDR strength functions calculated
with the force SLy6 in $^{238}$U. In both panels, the total strength (solid curve)
and its K=0 (dash curve) and K=1 (dotted curve) branches are depicted.}
\end{figure}

Here we discuss the anomalous deformation splitting of TDR in prolate $^{238}$U.
The corresponding SRPA and two-quasiparticle (2qp) strength functions are depicted
in Fig. 5. The strengths are smoothed by a Lorentz weight with the constant
averaging parameter $\Delta$= 1 MeV. The deformation parameter $\beta$=0.286
is obtained by minimization of the total energy of the system. Note that
the calculated $\beta$ excellently reproduces the experimental value
$\beta_{\rm{exp}}$=0.2863(24) \cite{Raman}.

Figure 5 shows that, in accordance with previous results for
rare-earth nuclei \cite{Kv14def}, the TDR in $^{238}$U also exhibits
the anomalous splitting. Indeed, in RPA case, the main K=1 peak lies
significantly lower (5-10 MeV) than the main K=0 peak (10-17 MeV). In
the low-energy region 5-12 MeV, the K=1 strength strongly dominates
over K=0 one. Though both branches are distributed in a wide energy
interval and have more or less the same strength at E$>$ 12 MeV, it is
easy to see that the K=1 centroid is certainly lower than the K=0
one. As seen from Fig. 5a), a similar picture emerges for the
unperturbed (without residual interaction) two-quasiparticle (2qp)
strength.

The anomalous splitting of the TDR might be related with the
fact that TDR, unlike the regular $E\lambda$ giant resonances
determined by $r^{\lambda}Y_{\lambda\mu}$-fields, is not the Tassie
mode. Following (\ref{tor_rel_1}), the toroidal multipole field rather
includes $r^{\lambda+2}Y_{\lambda\mu}$.  This means that simple
arguments explaining the deformation splitting in the GDR do not work
here. However, following expression (\ref{CM_divj1}), the CDR is also
not the Tassie mode.  At the same time, the calculations do not show
in the CDR the distinctive "opposite order" splitting in deformed
nuclei \cite{Kv13Sm}. Perhaps here the vortical character of the TDR
is a decisive factor.  This problem deserves further
exploration. Anyway, the anomalous splitting of the TDR can be used
for its experimental discrimination from other dipole modes.

\section{Summary}

Some remarkable properties of the isoscalar toroidal dipole resonance
(TDR), recently studied by our group, were briefly discussed. The main
attention was paid to a) the relation of the TDR and low-energy
  dipole strength (also denoted as PDR) \cite{Repko13}, b) the
possibility to use the toroidal flow as a measure of the nuclear
dipole vorticity \cite{Rein14vor}, and c) anomalous deformation
splitting of the TDR \cite{Kv14def}. The analysis was based on the
calculations within the self-consistent quasiparticle
random-phase-approximation (QRPA) method \cite{Ri80} with the Skyrme
force SLy6 \cite{SLy6}. Two QRPA versions, exact RPA for spherical
nuclei \cite{Repko} and separable RPA (SRPA) for deformed nuclei
\cite{Ne06} were used. As relevant examples, the spherical $^{132}$Sn
and deformed $^{238}$U were considered. The vortical TDR was discussed
together with its irrotational counterpart, the compression dipole
resonance (CDR).

Our study has confirmed previous findings
\cite{Kv11,Repko13,Rein14vor,Kv14def}. Namely, for the case of
$^{132}$Sn, we showed that the TDR and PDR share the same energy
region.  The field of the nuclear convection current in this region is
clearly toroidal. Both TDR and PDR flows have much in common at the
nuclear surface. So it is quite possible that the PDR is actually a
local manifestation of the TDR at the nuclear boundary.

Besides we presented the arguments that the familiar measure of the
nuclear vorticity proposed by Ravenhall and Wambach \cite{Ra87}
($j_+$ component of the nuclear current) is not relevant and obviously
fails in the TDR/CDR case.  Instead, the toroidal current and strength
are much better suited for this aim.

Finally, using the SRPA results for $^{238}$U, we considered the
anomalous deformation splitting of the TDR. In accordance to previous
study \cite{Kv14def}, we have found that in prolate $^{238}$U the
K=1 branch of the TDR has lower energy than the K=0 branch, i.e. we
received the opposite order of the branches as compared to the regular
giant dipole resonance. The nature of this feature is not yet clear
and needs a further study.  Anyway this feature may be used for an
experimental discrimination of the TDR.

Note that for dipole excitations in the PDR region the coupling with
complex configurations (CCC) can be important
\cite{End10,Li08,Ar12}. However, we think that the toroidal features
we have found are too strong to be spoiled by this effect. This is
confirmed e.g. by QPM calculations which include the CCC \cite{Ry02}
but nevertheless report a strong toroidal flow at the PDR region.

The isoscalar TDR and CDR dominate the E1(T=0) channel and can be
observed in $(\alpha, \alpha')$ reaction, see
e.g. \cite{Uchida04}. So, despite of their second-order character
\cite{Kv11}, the PDR and CDR are accessible and important examples of
the dipole excitations. The TDR is the only vortical mode in E1(T=0)
channel. The TDR and PDR share the same energy region and are
certainly related. Now the PDR is intensively investigated as it
provides an important information for the nuclear equation of state
and astrophysical applications \cite{Paar07,Savran13}. The analysis of
the TDR/PDR interplay can be essential for a better understanding of
the PDR features and related topics.

\section*{Acknowledgments}
The work was partly supported by DFG (RE 322/14-1), Heisenberg-Landau
(Germany-BLTP JINR), and Votruba - Blokhintsev
(Czech Republic - BLTP JINR) grants.
J.K. and A.R. are grateful for the support
of the Czech Science Foundation (P203-13-07117S).

\end{document}